\documentclass[preprint,proceedings]{rmaa}
\usepackage{paralist}
\usepackage{psfrag,color}

\SetYear{2004}
\SetConfTitle{Proceedings of the II International Workshop on Science with GTC}
\title{Searching for High Redshift Clusters} 
\author{
  R. Juncosa\altaffilmark{1},
  C.M. Guti\'errez\altaffilmark{1},
  and A. Fern\'andez-Soto\altaffilmark{2}}
\altaffiltext{1}{Instituto de Astrof\'\i sica de Canarias, La Laguna, Tenerife, Spain.}
\altaffiltext{2}{Observatori Astron\`omic, Universitat de Val\`encia, Spain.}
\shortauthor{Juncosa, Guti\'errez, \& Fern\'andez-Soto}
\fulladdresses{
\item R. Juncosa and C.M. Guti\'errez: Instituto de Astrof\'\i sica de Canarias, La Laguna, Tenerife,
Spain (\email{robert@iac.es, cgc@iac.es}).
\item A. Fern\'andez-Soto: Observatori Astron\`omic, Universitat de Val\`encia,
Burjassot, Val\`encia, Spain (\email{alberto.fernandez@uv.es}).
}
\listofauthors{R. Juncosa, C. M. Guti\'errez, \A. Fern\'andez-Soto}
\indexauthor{Juncosa, R.}
\indexauthor{Guti\'errez, C. M.}
\indexauthor{Fern\'andez-Soto, A.}
\abstract{
Distant clusters of galaxies provide a powerful method to study the
formation and evolution of galaxies, and large scale structure of the
Universe. However, the number of known clusters at high redshift ($z>
0.5$) is still very reduced. As a preparatory work for detailed studies
wigh GTC, we are searching for high redshift clusters in public wide
optical surveys. We will complement this with near IR observations in 4 m
telescopes. Here, we present some preliminary results. In a field of
$35\times 35$ arcmin$^2$ we have detected  8 clusters of galaxies at redshift  
$z> 0.5$.
}
\resumen{
Los c\'umulos de galaxias lejanos nos permiten estudiar
la formaci\'on y la evoluci\'on de las galaxias y de la estructura del
Universo a gran escala. Sin embargo, el  n\'umero conocido de c\'umulos
con alto corrimiento al rojo ($z> 0.5$) es muy reducido. Como
preparaci\'on para estudios detallados  con GTC, estamos tratando de
localizar un n\'umero suficiente de estos c\'umulos a alto $z$ en
cartografiados de gran campo  en bandas \'opticas, que complementaremos
con observaciones en el infrarrojo cercano con telescopios  de 4 m.
Aqu\'\i~presentamos algunos resultados preliminares.  En un campo de
35x35 minutos de arco cuadrados hemos detectado 8  c\'umulos de galaxias con $z>
0.5$.
}
\addkeyword{clusters of galaxies}
\addkeyword{galaxy evolution}
\addkeyword{galaxy formation}
\addkeyword{cosmology}
\begin{document}
\maketitle
\section{Introduction}
\label{sec:intro}

The study of high redshift clusters ($z>0.5$) is an important tool to
understand both galaxy and cluster formation and evolution. For instance its
number and distribution in mass and redshift  allow to discriminate between
different cosmological scenarios. Fig. 1 presents the expected density of
clusters for a standard cosmological lambda cold dark matter model. Although
the number of known high redshift clusters has increased in the last years,
this number is still reduced and makes difficult statistical studies. The
detection of  representative numbers ($\sim 50-100$) of such clusters would 
require the analysis of deep images covering several square degrees. Appropriate
surveys in the optical exist now and will be available in the near future at
near infrared wavelengths. In this contribution, we summarize our work in this
area, and presents the analysis of a region of the sky $35\times 35$ arcmin$^2$
in several optical filters in which we have detected 8 candidates to be clusters 
of galaxies at $z>0.5$. Additional details are given in [2]

\begin{figure}[!t]
  \includegraphics[width=\columnwidth]{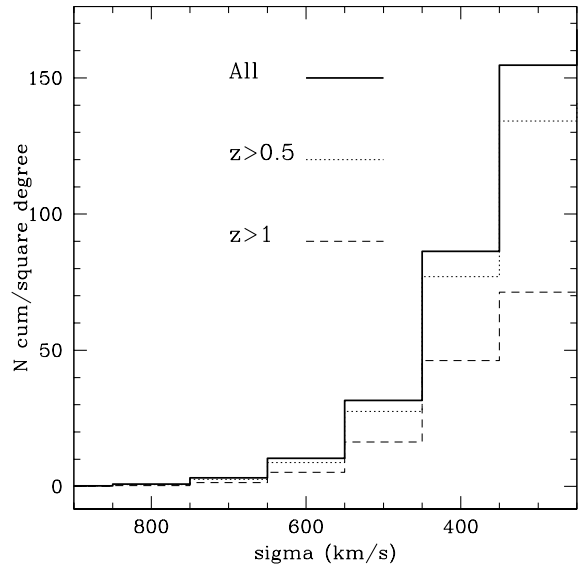}
  \caption{Expected density of clusters and galaxy groups as a function of the
  velociy dispersion (or equivalently mass). The diagrams are
  based on simulations by the Virgo Consortium [1] assuming a flat model 
  with $\Omega_{CDM}= 0.3$ and $\Omega_\Lambda= 0.7$.}
  \label{fig:simple}
\end{figure}

\section{Data}
\label{sec:errors}

We are analyzing the NOAO Deep Survey [3] and the Deep Lens Survey [4]. The
NOAO Deep Survey covers an area of 18 square degrees  in the $B_w$, $R$, $I$,
$J$, $H$, $K$ filters  with 5$\sigma$ limiting magnitude $R\sim 25$ mag.
So far, 1 square degree in the optical have been released. The Deep Lens
Survey has an area of 27 square degrees (6 of them are public now), in the
$B$, $V$, $R$ and $z^\prime$ filters  (some fields have been observed also in
the $I$ band). The limiting magnitude is $R\sim$ 26.5 mag.

\section{Analysis and results}
\label{sec:EPS}

The results presented here have been obtained determining the
photometric redshifts [5] and looking for overdensities in the ($RA$,
$Dec.$, $redshift$) space.   So far, we have analyzed a region of $35\times 35$ arcmin$^2$ in which
we have found several  likely candidates
to be high redshift clusters. Fig.~2 shows the field analyzed and the
positions of the cluster candidadates at $z>0.5$. A list with the estimated redshifts, number
of members detected and the significance of each cluster candidate is 
presented in Table~1. The color-magnitude ($V-I$ $vs.$ $I$) diagram and the distribution in
redshift in one of the regions in which a cluster candidate is present
are shown in Figs. 3 and 4 respectively.
\begin{figure}[!t]
  \includegraphics[width=\columnwidth]{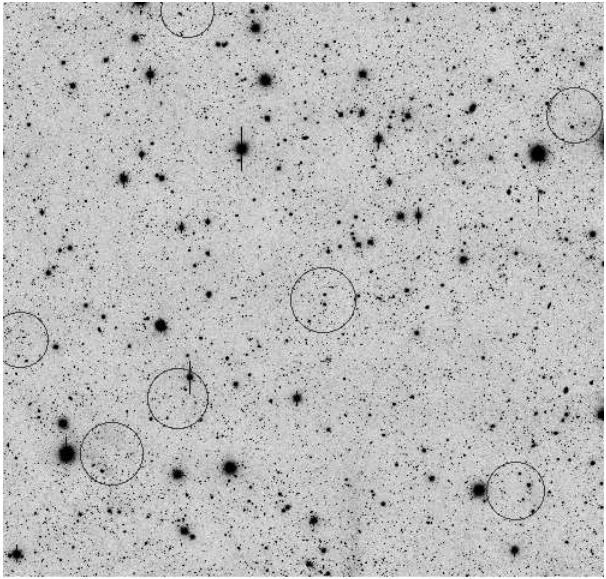}
  \caption{Deep Lens Survey image in the R filter covering an area of 35x35 
  arcmin. The circles enclosed the position where cluster candidates at $z>0.5$
  have been found.}
  \label{fig:simple}
\end{figure}

\begin{table}[!t]\centering
  \setlength{\tabnotewidth}{\columnwidth}
  \tablecols{3}
  \setlength{\tabcolsep}{1.0\tabcolsep}
  \caption{Cluster candidates at $z>0.5$}\label{tab:ion_ab}
  \begin{tabular}{ccc}
    \toprule
    Redshift & Num. obj. zone & Reliability \\
    \midrule
  $0.90$ & $6$ & $>0.99$\\
  $0.85$ & $5$ & $ 0.98$\\
  $0.55$ & $6$ & $ 0.90$\\
  $0.60$ & $5$ & $ 0.86$\\
  $0.80$ & $4$ & $ 0.82$\\
  $1.50$ & $3$ & $ 0.82$\\
  $1.40$ & $3$ & $ 0.80$\\
      \bottomrule
    \tabnotetext{}{Table 1. Candidates to be clusters of galaxies 
    at redshift $>0.5$ in
    one of the fields of the Deep Lens Survey.}
  \end{tabular}
\end{table}

\begin{figure}[!t]
  \includegraphics[width=\columnwidth,height=4.8cm]{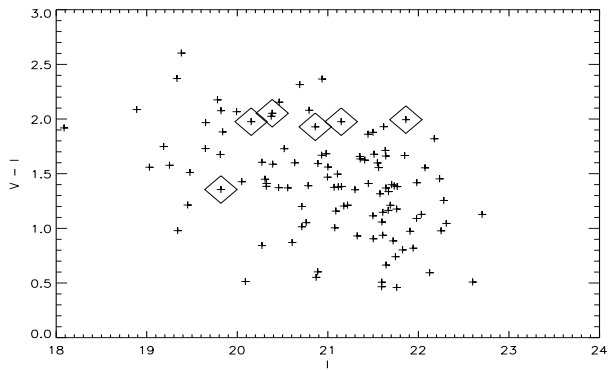}
  \caption{Color-magnitude diagram of all the galaxies ({\it crosses}) enclosed in one of the circles shown in Figure~2; objects
  identified as cluster members galaxies ({\it diamonds}) follow a
  tight $V-I$ $vs.$ $I$ relation.
  }
  \label{fig:simple}
\end{figure}

\begin{figure}[!t]
  \includegraphics[width=\columnwidth,height=4.8cm]{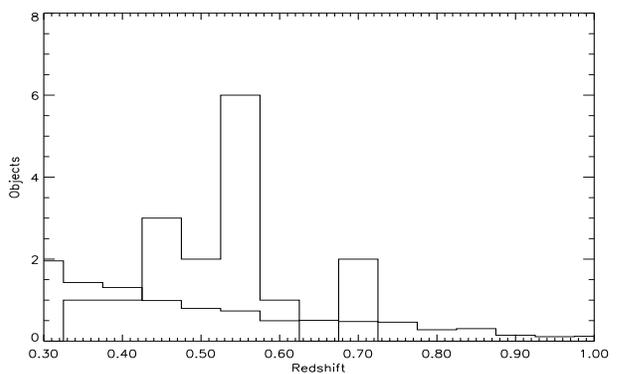}
  \caption{The figure shows as a function of redshift: ({\it dotted})
  the mean number of objects in a region of $2\times 2$  arcmin$^2$ 
  of the Deep Lens
  Survet field shown in Fig.~2. ({\it Dashed}) the expected 1$\sigma$ 
  upper limit for a random $poisson$
  distribution, and ({\it solid}) the actual distribution of
  objects in such region. A clear overdensity exists at $z\sim 0.6$. }
  \label{fig:simple}
\end{figure}


\end{document}